\documentclass[twocolumn,preprintnumbers,amsmath,amssymb]{revtex4}
\usepackage{graphicx}
\usepackage{dcolumn}
\usepackage{bm}
\begin{document}
\title{\bf Lattice thermal properties of  Graphane: thermal contraction, roughness and heat capacity}
\author{M. Neek-Amal$^1$ and F. M. Peeters$^2$ }
\affiliation{$^1$Department of Physics, Shahid Rajaee University,
Lavizan, Tehran 16785-136, Iran.\\$^2$Departement Fysica,
Universiteit Antwerpen, Groenenborgerlaan 171, B-2020 Antwerpen,
 Belgium.}
\date{\today}
\begin{abstract}
Using atomistic simulations we determine the roughness and the thermal properties of a suspended graphane sheet.
As compared to graphene we found that hydrogenated graphene has: 1) a larger thermal contraction,
2) the roughness exponent at
room temperature is smaller, i.e. $\simeq$ 1.0 versus $\simeq$ 1.2 for graphene, 3) the wave lengths of
the induced ripples in
graphane cover a wide range corresponding to length scales in the range (30-125)\,\AA~at room temperature, and 4)
the heat capacity of graphane is estimated to be 29.32$\pm$0.23\,J/molK
 which is $14.8\%$ larger than the one for graphene, i.e.
 24.98$\pm$0.14\,J/molK. Above 1500\,K we found that  graphane
 buckles when its edges are supported in the $x-y$ plane.
\end{abstract}
\maketitle
\section{Introduction}
Graphane (GA), a two dimensional covalently bonded hydrocarbon, was
first predicted from ab-initio calculation  by Sluiter~\emph{et
al}~\cite{sluiter} and recently rediscovered~\cite{sofo}. In a
recent experiment Elias~\emph{et al}~\cite{elias} demonstrated the
fabrication of GA from  a graphene (GE) membrane through
hydrogenation which was found to be reversible. Density functional
theory and molecular dynamics simulations
 employing different force fields were carried out to study the
structural and electronic properties of both GA and
GE~\cite{acs,peeters}.

Hydrogen (H) atoms are chemically bound to the carbon (C) atoms on
 alternating sides of the membrane  (chair like conformer) which
causes a local buckling of the membrane. Such deformations for small
membrane sizes ($<1 nm$) have been recently reported~\cite{Hfrust}.
In the early stages of the hydrogenation process,
membrane shrinking and extensive membrane corrugations occur due to
 the formation of a significant percentage of
uncorrelated H frustrated domains~\cite{Hfrust}.

The morphology of perfect GA,  and its comparison with a perfect GE
membrane, has  not yet been investigated for large samples.
Ab-initio calculations are restricted to small unit cells and
therefore we will use atomistic simulations to show the main
differences between the morphology of a large sample of GA and GE.
Our atomistic simulations are based on the second generation of
Brenner potential~\cite{brenner2002} (REBO) which includes the
interaction with third nearest neighbors via a bond-order term that
is associated with the dihedral angles. Therefore, such a potential
is suitable for various atomistic simulation purposes including the
calculations the lattice thermal properties (phonon
dispersion~\cite{phononBrenner}, elastic moduli~\cite{Lu2009},
thermal conductivity~\cite{brennerthermal}, etc) of carbon
nanotubes, graphene and  hydrocarbons. Nevertheless, there are a few
shortcomings in REBO potential which are important when modeling
processes involving energetic atomic collisions (because both
Morse-type terms approach finite values at very small atomic
distance), and the neglect of a separate $\pi$ bond contribution
which leads to the overbinding of radicals our study is insensitive
for these shortcomings). Of course, the electronic properties of GE
and GA and the thermodynamical properties at low temperatures
(because of the e.g. quantum zero energy) are beyond REBO's ability.
In those situations ab-initio calculations, particularly those based
on density functional theory (DFT), are extremely useful. Ab-initio
molecular dynamics simulation (e.g. Car-Parrinello molecular
dynamics~\cite{carpar}) is a highly appreciated theory which
eliminates the force-fields based restrictions, but its main
disadvantage is that it is only applicable for small size systems
(typically $N<$ 500 atoms).

Here we study relatively large systems and consider thermal and
structural properties of GE and GA above 50\,K and show that for a
GA sample with size 183~\AA$\times$185~\AA~containing 22400 atoms,
the roughness, the induced ripples structure and the total energies
are very different from  a GE membrane. It has been shown that the
ripples in GE strongly affect its thermo-electronic
properties~\cite{ripples}. Temperature effects are studied and we
found that the thermal contraction of a suspended GA is larger than
the one for GE. Moreover the calculated roughness exponent indicates
that GA is more rough than GE even at room temperature. From our
simulations, we predict that the ripples in GA affect its
thermo-electronic
 properties much more than for GE.

This paper is organized as follows. In Sec.\,II  we introduce the
atomistic model and the simulation method. Section\,III contains our
main results for both graphene and graphane.
 Results for thermal contraction, roughness and heat capacity for different temperatures
are presented and compared to available experimental results. In
Sec.\,IV we conclude the paper.

\begin{figure*}
\begin{center}
\includegraphics[width=1.0\linewidth]{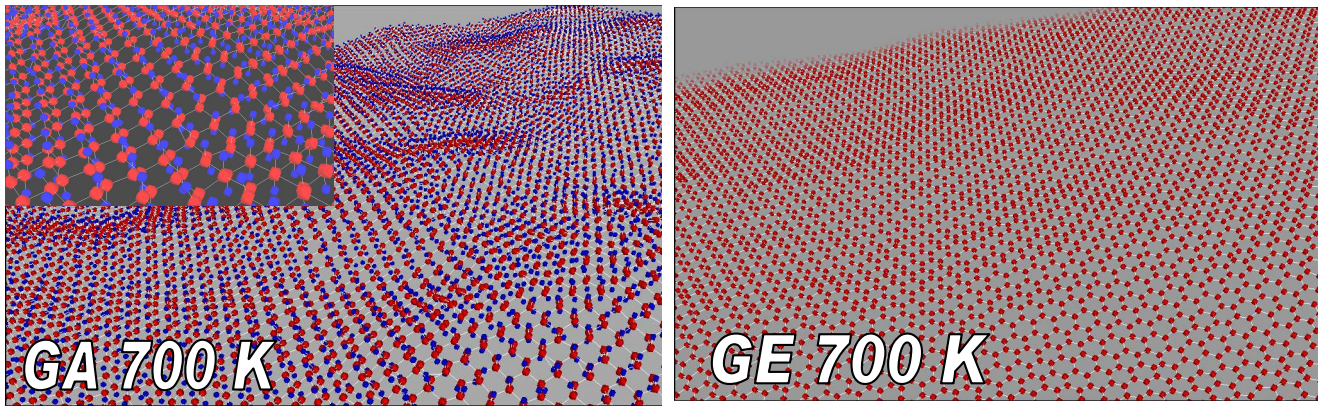}
\caption{(Color online) Two snap-shots of suspended graphane (GA)
and graphene (GA) at 700\,K. The inset is a zoomed region which
shows a regular alignment of C-H bonds.\label{figmodel} }
\end{center}
\end{figure*}
\section{Method and model}
 Classical atomistic molecular dynamics
simulation (MD) is employed  to simulate large flakes of GE and GA
at various temperatures. The second generation of Brenner's
bond-order potential is employed which is able to describe covalent
sp$^3$ bond breaking and the formation of associated changes in
atomic hybridization within a classical
potential~\cite{brenner2002}. The Brenner potential (REBO) terms
were taken as $E_P=\sum_i\sum_{j>i}[ V^R(_{ij})-B_{ij}V^A(r_{ij}]$,
where $E_P$ is the average binding energy, and $V^R$ and $V^A$ are
the repulsive and  attractive term, respectively, where $r_{ij}$ is
the distance between atoms \emph{i} and \emph{j}. $B_{ij}$ is called
the bond order factor which includes all many-body effects. $B_{ij}$
depends on the local environment of the bond i.e. the bond and
torsional angles, the bond lengths and the atomic coordination in
the vicinity of the bond. This feature allows the Brenner potential
to predict correctly the configurations  and energies for many
different hydrocarbon structures and the correct hybridization.

 Initially the coordinates of all carbon atoms in GA (GE) are put
in a flat surface of a honey comb lattice with nearest neighbor
distance equal to $a_0=0.153$~nm (0.142~nm). The hydrogen
 atoms are alternatively put on top and bottom of  the carbons, i.e.
 $(CH)_n$ with $n=11200$ (chair like model for
 GA~\cite{sofo}, see the inset in the left panel of  Fig.~\ref{figmodel}). To make sure that the second
generation of the Brenner potential gives the correct force  field
for our thermo-mechanical studies of the GA sheet, we performed
extra simulations  where we started with a non optimum bond length
of $a_0=0.142$\,nm. Already after only a few time steps the correct
optimum GA lattice spacing, i.e. $a_0\sim$0.153~nm, was found
confirming the ability of the used potential for GA simulations.

The considered systems are square sheets with dimensions
183~\AA$\times$185~\AA~for GA and 170~\AA$\times$170~\AA~for GE, in $x$ and $y$- directions where we
considered both armchair and zigzag edges.
 The number of atoms in GA (GE) is $n=22400$ ($n$/2).
 We simulated the system at non-zero temperatures (above 50\,K)  by
employing a Nos'e-Hoover thermostat. In order to mimic the
experimental set-up we prevent  motion along the $z$-direction at
the ends of the system along $x$. This is realized in practice by
fixing a row of atoms at both longitudinal sides. This  supported
boundary condition prevents the drift (of the ends of the sheet) in
the $z$-direction. Motion in the $x-y$ plane is allowed which allows
the system  to relax and to exhibit lattice contraction/expansion.

Before starting the sampling, we let the system to find its true
size and temperature during the first 5\,ps. During this process GA
and GE change their longitudinal length so that we always observe
lattice contraction implying that GA and GE shrink through surface
corrugation (i.e. creating ripples)~\cite{PRB821}. Figure~\ref{figmodel} shows two snap-shots of GA and GE
relaxed at 700 K. The C-H bonds (except at the boundaries) are
almost parallel which is a consequence of lateral H-H repulsion.

\section{Results and discussion}
\subsection{Equilibrium length and thermal contraction} After
equilibrating, we start to record the longitudinal size (in $x-y$
plane along $x$-axis) of the system for 2\,ps. Averaging over the
lengths during the sampling process gives the average length of the
system, i.e. $\langle L\rangle$, at the given temperature.
Figure~\ref{fig1}(a) shows the  thermal contraction coefficient,
i.e. $\gamma_L=\frac{dL}{L_0dT}$, versus temperature,
 where $L_0=183\,\AA$~($L_0=170\,\AA$)~is the initial length of
 GA (GE) (for a flat sheet at $T$=0\,K) and $dL=\langle L\rangle-L_0$. As we
see $\gamma$ increases with  temperature and is always negative in
the studied temperature range. Surprisingly GA has a larger
contraction with temperature than GE.  These curves are fitted to
$\alpha+\beta/T$ where $\alpha$=22$\pm$5$\times$10$^{-6}$\,K$^{-1}$,
4$\pm$1$\times$10$^{-6}$\,K$^{-1}$ and $\beta$=-877.10$\pm$29 and
-298.8$\pm$6 for GA and GE, respectively (solid curves in
Fig.~2(a)).

At present a few studies exist on the negative thermal expansion
coefficient (TEC) of GE, but none are available for GA. The actual
values of the TECs for GE differ from one study to the other
depending on e.g. the size and the boundary
conditions~\cite{thesis}. Mounet and Marzari~\cite{marzari}
 presented a study of the thermodynamical and structural properties
of carbon based structures (graphite, diamond and GE) using GGA-DFT
calculations. In order to calculate the finite temperature thermal
expansion and heat capacity, a quasi harmonic approximation (QHA)
 was employed~\cite{marzari,thesis}.
Using the DFT results for the phonon dispersion relation and
minimizing the QHA free energy with respect to the lattice parameter
($\frac{\partial F}{\partial a}|_T=0$), they found the linear
thermal expansion of the lattice parameter -$a$- for graphite, GE
and diamond. The linear thermal expansion for the lattice parameter
of GE, $\gamma_a=\frac{1}{a}\frac{\partial a}{\partial T}$ where
$a=\sqrt{3}a_0$, was found to be negative and about
$\gamma_a\simeq-0.35\times10^{-5}\,K^{-1}$ at room temperature.
QHA+GGA includes only weakly anharmonicity but
 it includes quantum effects and zero point energy.
 The strong anharmonic coupling of the bending and stretching
 modes in GE, which  are essential for the GE
 stability, are accurately described  by classical empirical potentials at non-zero
  temperatures~\cite{fasolinonature}.
In our study the negative thermal contraction is reported for the
length -$L$- of GA and GE samples. Since ripples appear in GE and GA
membranes, we expect a contraction of the length of the systems.
$\gamma_L$ is not directly related to $\gamma_a$. In fact, $dL$ is
the variation of the measured length  along $x$-direction while $da$
is the variation of the lattice parameter not necessarily in the
$x-y$ plane. Therefore our results for the contraction of the GE's
length are not directly related to those of QHA+GGA for contraction
of the lattice parameter.

 In a recent
experiment~\cite{natnano} the morphology of different graphene
membranes (micron size membranes)
 suspended across trenches
on Si/SiO$_2$ substrates were investigated using AFM and STM. TECs
were found to be about $\gamma_{exp}\sim-0.7\times10^{-5}\,K^{-1}$
at room temperature. With increasing temperature the measured TECs
approaches
 zero nonlinearly. Here, we find a similar contraction in the longitudinal size
 (e.g. $\gamma_L\sim-9.0\times10^{-5}\,K^{-1}$ at $T$=300\,K
for GE)
 which approaches  zero with increasing
temperature but which is an order of magnitude larger. To compute
the thermal contraction we calculated the longitudinal size (not the
arc length as done in the experiment~\cite{natnano}). There are
important differences between our approach  and those from the above
mentioned experiment. First, we use the initial length (as a
reference for the length) of our samples as the length of the sheet
at zero temperature. Second, the sample size is different. Third,
there is no  substrate in our study which could reduce the
contraction. Our suspended sample is free to drift in the $x-y$
plane while in the experiment, the substrate prevents free drift in
the $x-y$ plane in addition to movement along the $z$-direction
which causes a lower contraction. Notice that
$\gamma_L<\gamma_{exp}<\gamma_a$.
 We found that in  thermal
equilibrium and under  supported boundary conditions, the
equilibrium size of GE is longer than GA. Therefore, the GA surface is
much more corrugated than GE.

The larger contraction of GA is due to the larger amplitude of the
ripples as compared to GE. The reason is that hydrogen atoms which
are below and above the sheet attract or repel the carbon atoms and
push them in different directions, (i.e. random thermal fluctuations
of the hydrogens at finite temperature). Therefore, we expect an
increasing randomness of the GA sheet and expect the formation of
different patterns of ripples in a GA sheet and a larger
corrugation. The thickness of GA is larger than GE, hence one
expects for such a thicker material, ordinary positive thermal
expansion (or at least smaller thermal contraction with respect to
GE). However, we found negative thermal contraction for GA which is
larger than the one for GE. It is interesting to note that above
1500\,K we observe a buckling of GA. Figure~\ref{fig2000} shows a
snap shot of a buckled GA sheet at 2000\,K. A similar buckling
 was observed in experiments on suspended GE~\cite{natnano}
and in a circular GE sheet subjected to radial
strain~\cite{neekJPCM}. Therefore, increasing temperature induces  larger
axial strains in GA.

\subsection{Roughness} 
\begin{figure*}
\begin{center}
\includegraphics[width=0.325\linewidth]{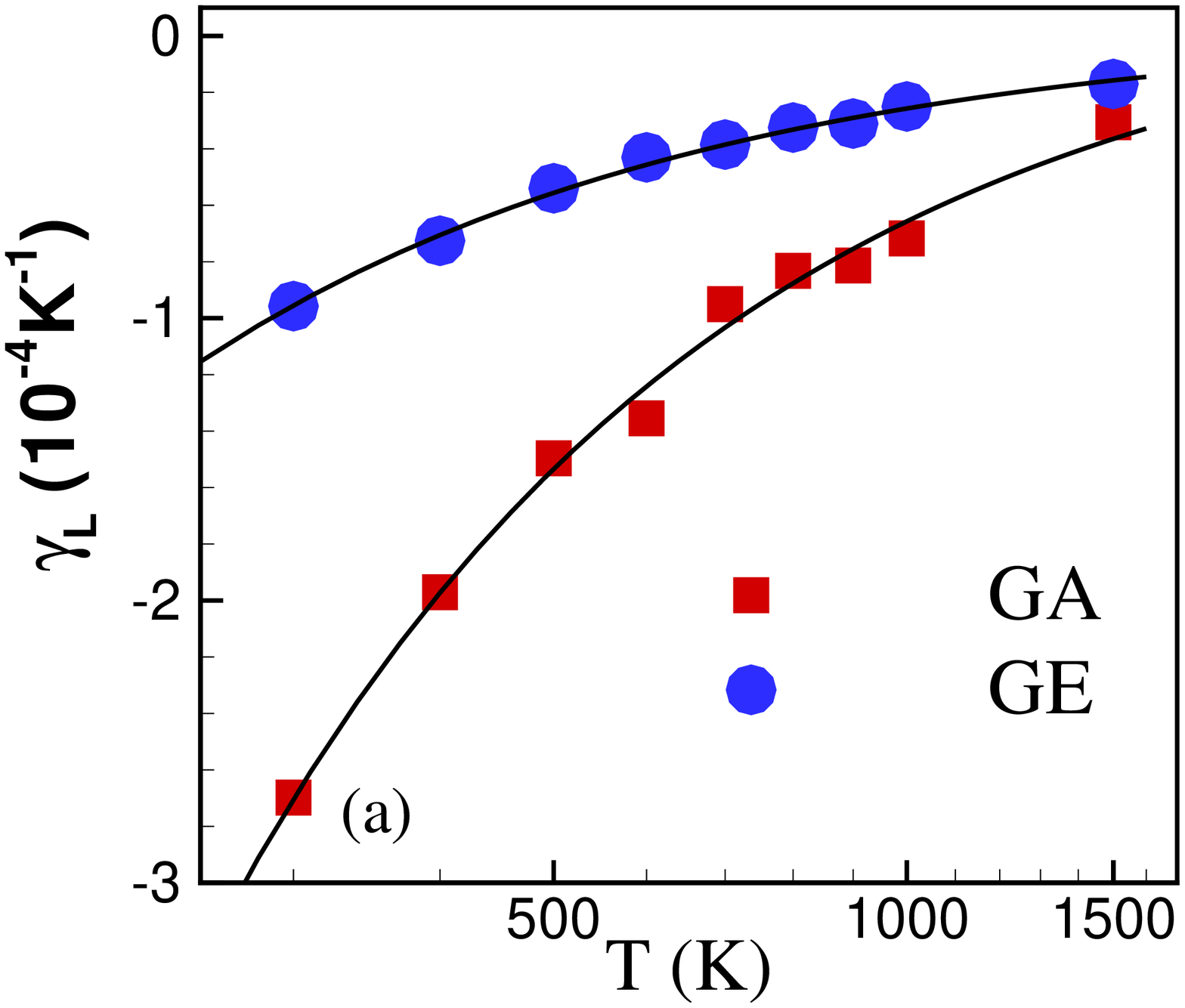}
\includegraphics[width=0.325\linewidth]{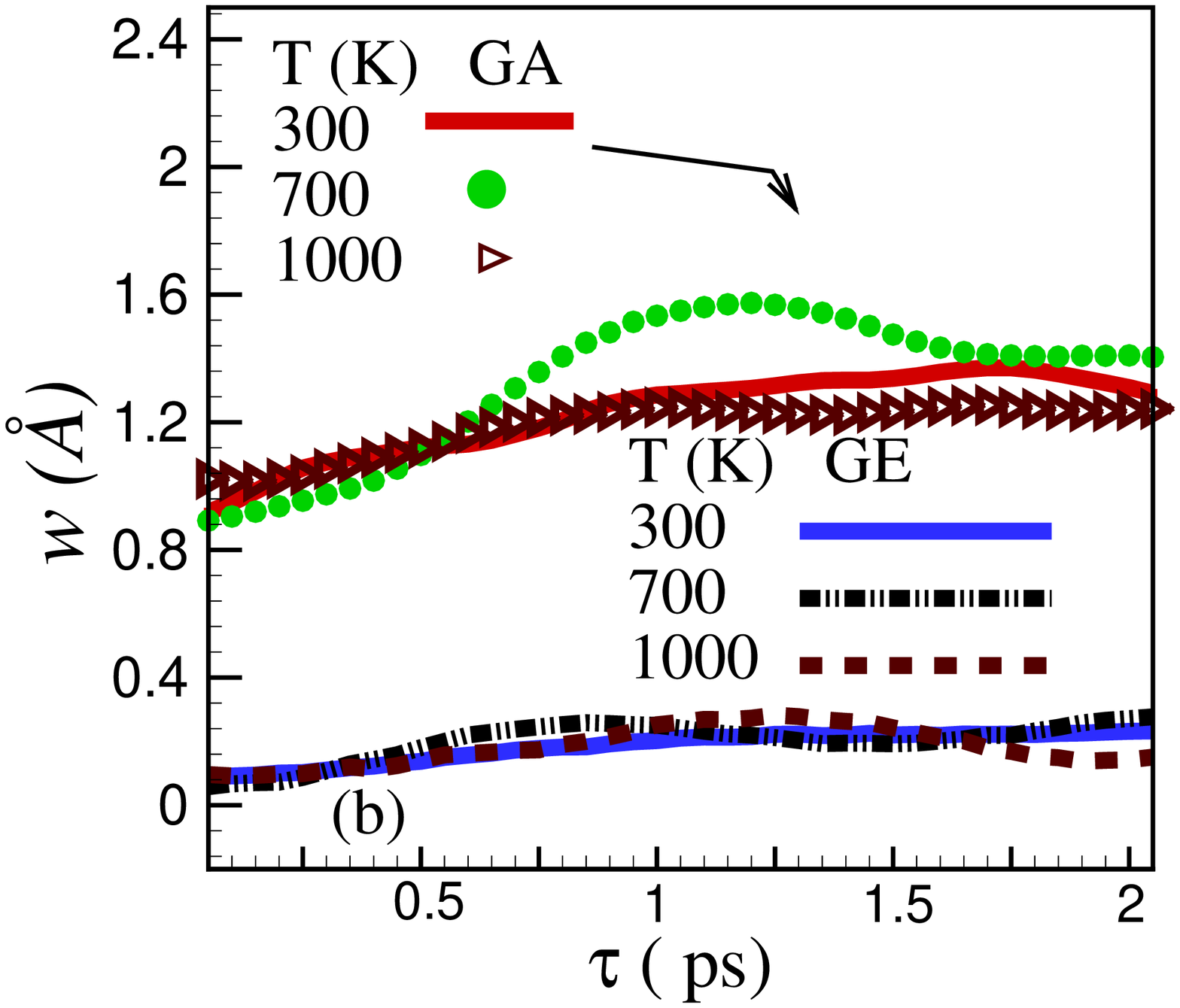}
\includegraphics[width=0.325\linewidth]{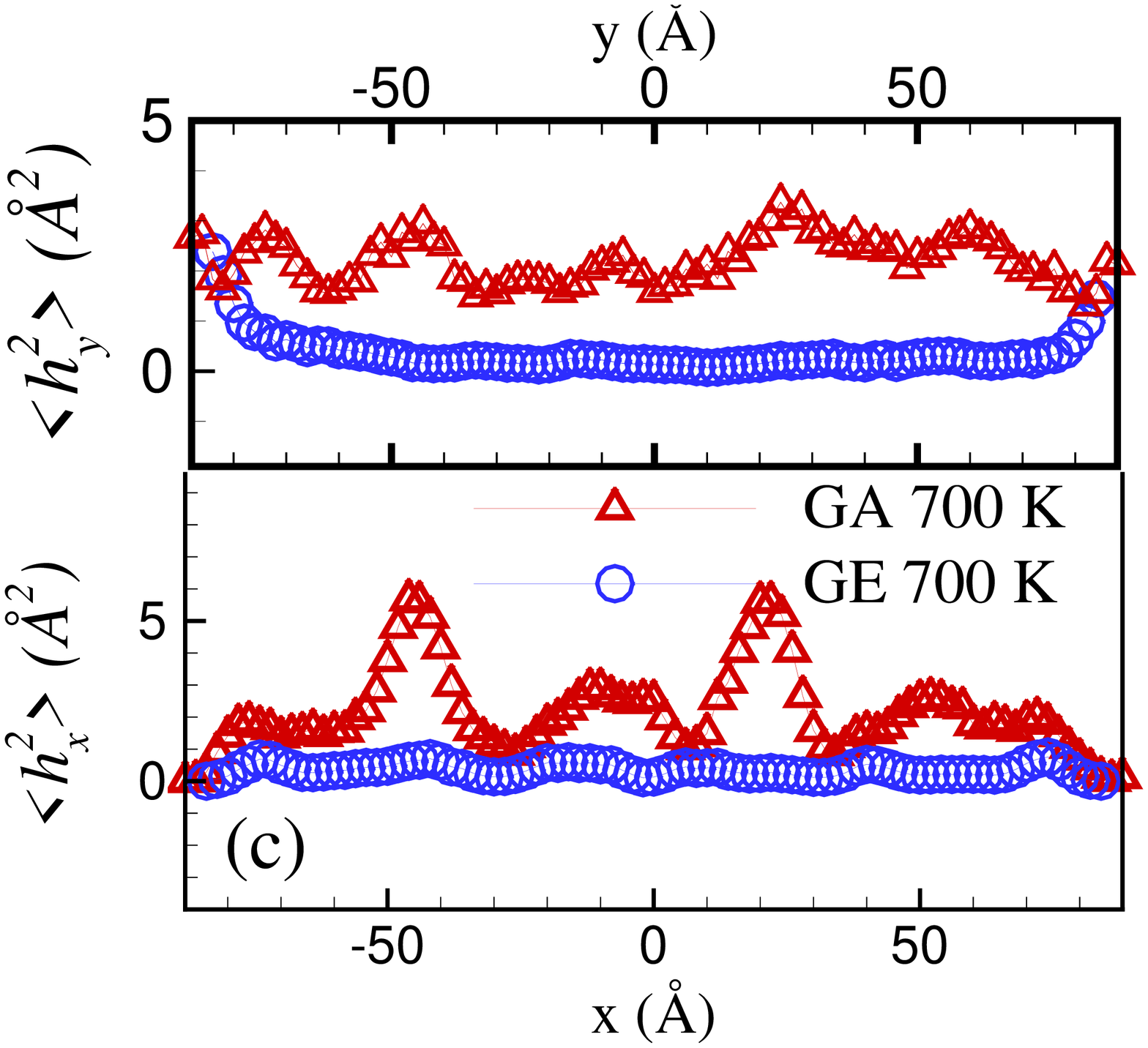}
\caption{(Color online) (a) Temperature dependence of the thermal
contraction coefficient of a square sheet of GA and GE subjected to
longitudinal supported boundary  and lateral free boundary
conditions. Solid curves are fits according to $\alpha+\beta/T$. (b)
Roughness of GA and GE as a function of time. (c) The variation of
$\langle{h^2_x}\rangle$ (averaged over lateral size) and
$\langle{h^2_y}\rangle$ (averaged over longitudinal size) at
700\,K.\label{fig1} }
\end{center}
\end{figure*}

\begin{figure}
\begin{center}
\includegraphics[width=1.0\linewidth]{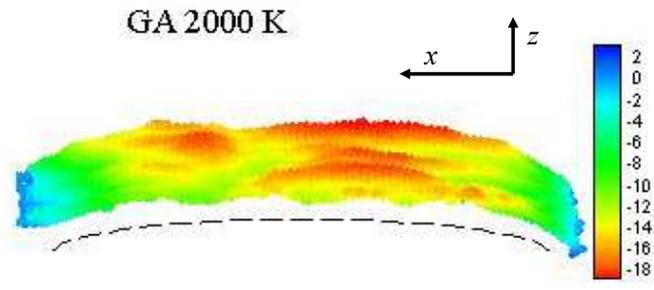}
\caption{(Color online) Buckling of graphane at 2000\,K.
Dashed curve shows the convex shape of  graphane.}
\label{fig2000}
\end{center}
\end{figure}

\begin{figure}
\begin{center}
\includegraphics[width=0.42\linewidth]{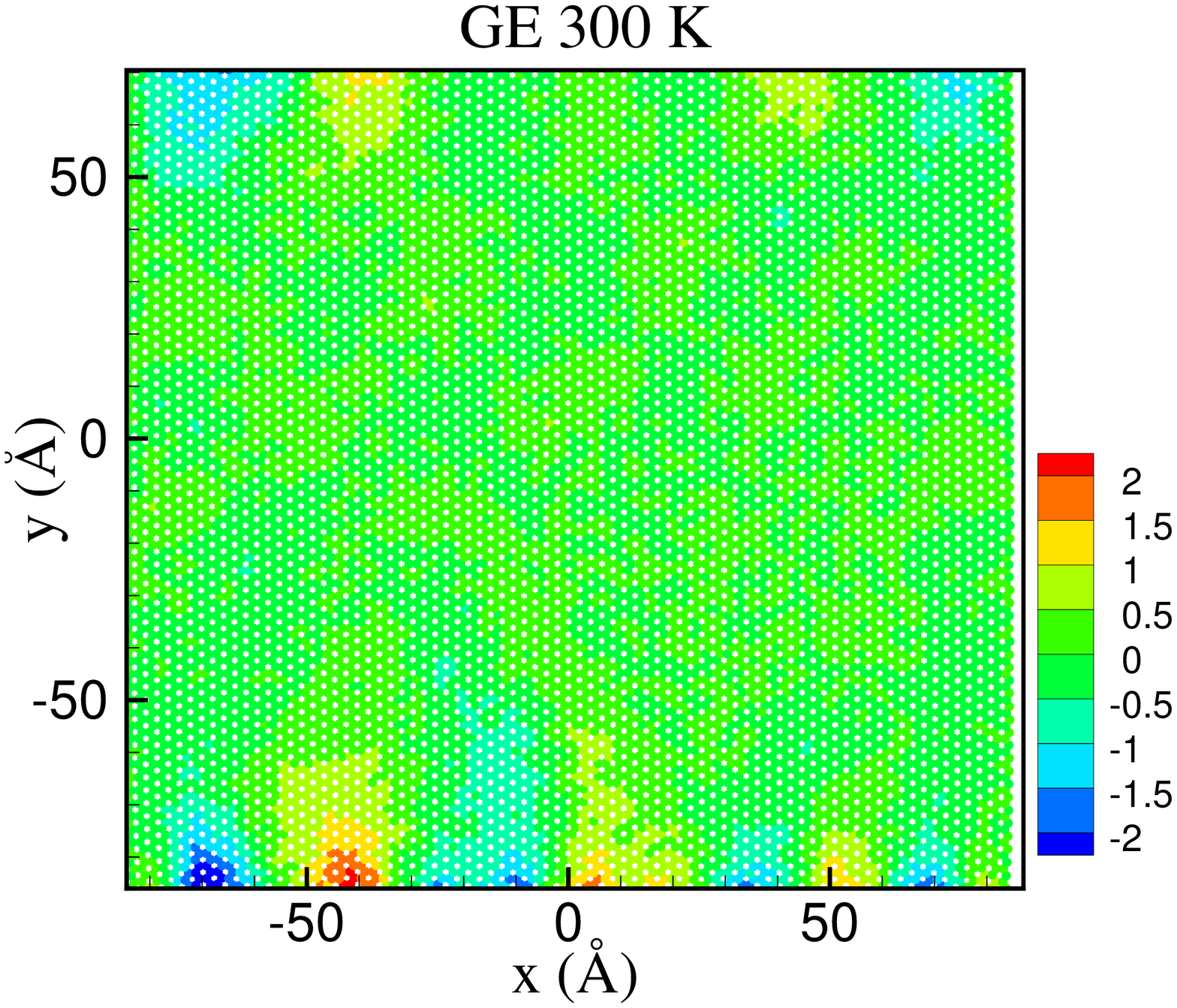}
\includegraphics[width=0.42\linewidth]{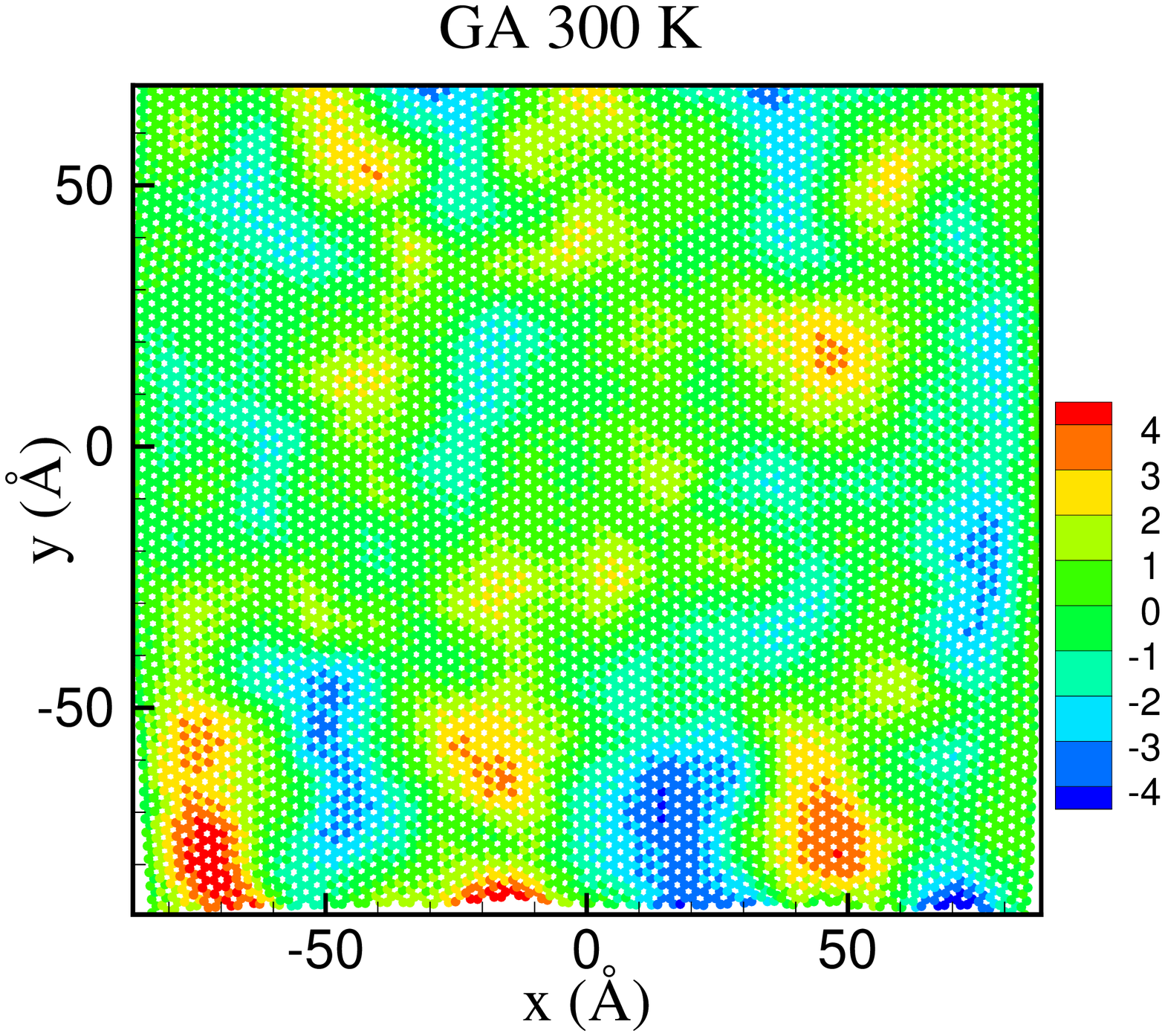}
\includegraphics[width=0.42\linewidth]{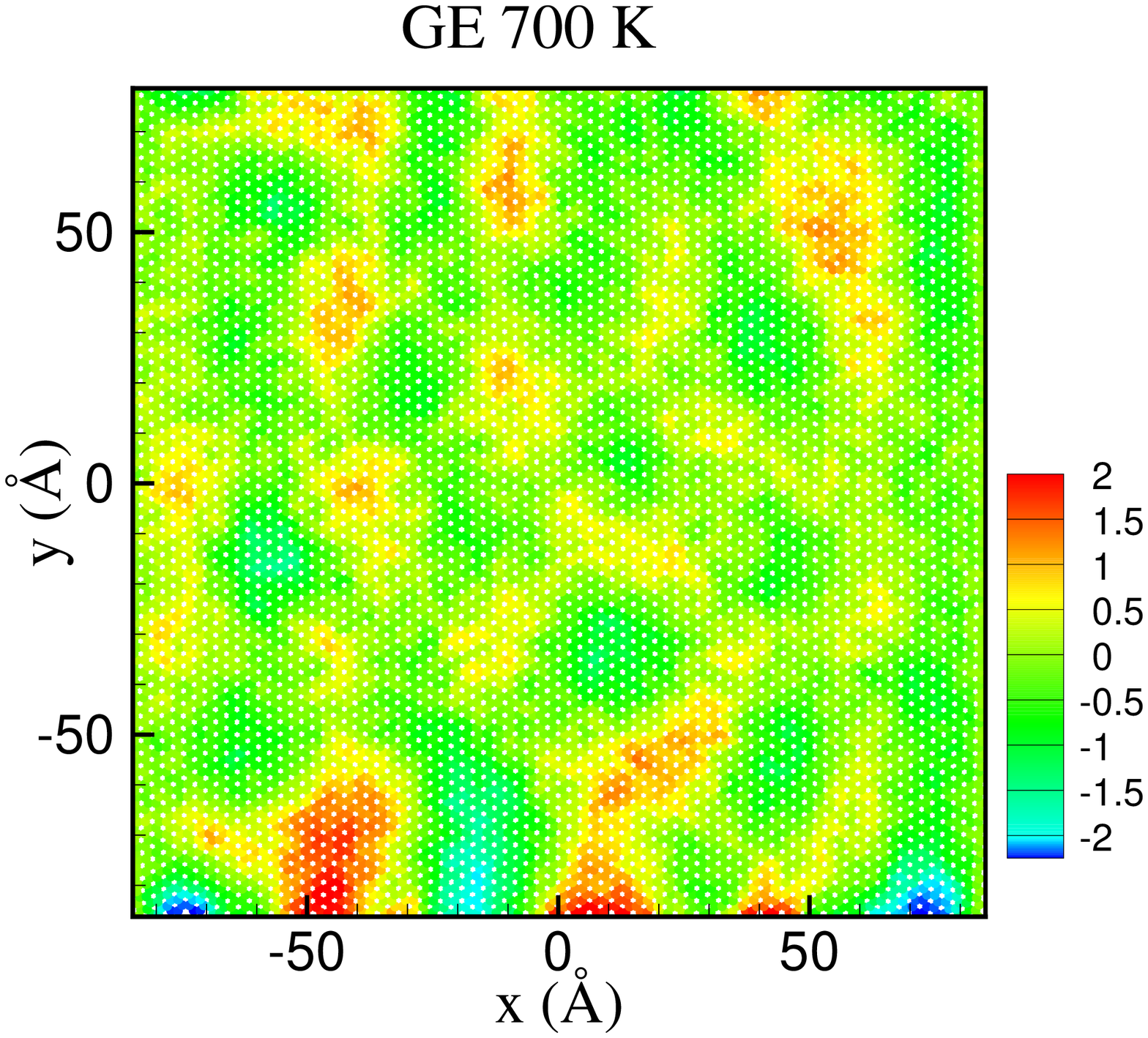}
\includegraphics[width=0.42\linewidth]{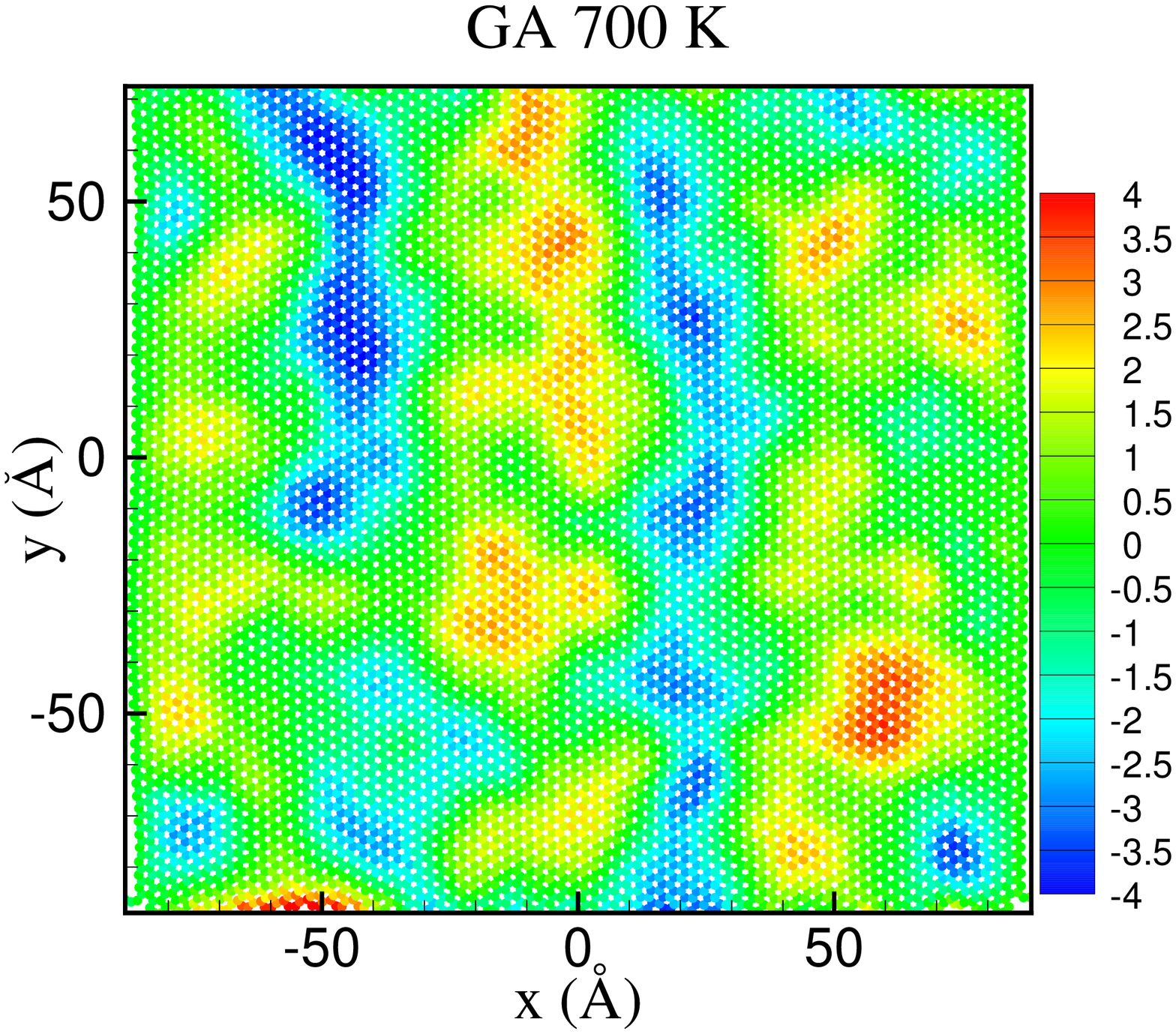}
\caption{(Color online) Contour plot of the $z$-position of the carbon atoms of GA and GE
at 300\,K and 700\,K corresponding to the situation of
Fig.~\ref{fig1}(c). The lateral edges are free while the longitudinal
edges are not allowed to move in the $z$-direction. \label{fig2D} }
\end{center}
\end{figure}

\begin{figure}
\begin{center}
\includegraphics[width=0.8\linewidth]{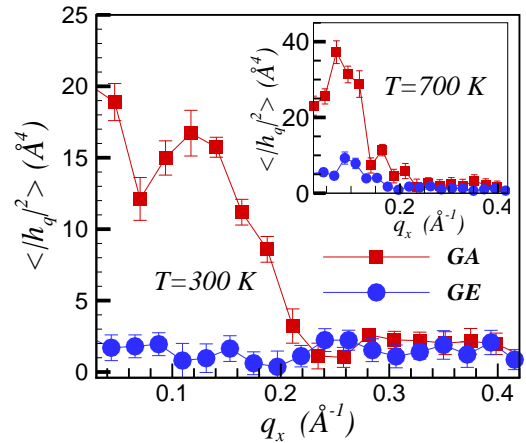}
\caption{(Color online) The amplitude of the Fourier transform of
$h_x$ of GA and GE atoms along the $x$-direction
 (averaged over 60 samples) for two temperatures T\,=\,300\,K and T\,=\,700\,K (the inset).\label{figFourier} }
\end{center}
\end{figure}

Figure~\ref{fig1}(b) shows the average square
root out of plane deviation of the carbon atoms (i.e. static
roughness, i.e. $w=\sqrt{\langle h^2\rangle-\langle h\rangle^2}$)
versus time for three typical temperatures. For long times $w$
fluctuates around 0.2\,\AA~for GE and 1.3\,\AA~ for GA which is 6.5
times larger.
 Figure~\ref{fig1}(c) shows the variation of
$\langle{h^2_x}\rangle$ (where average is over lateral size of the
system) and $\langle{h^2_y}\rangle$ (where average is  over
longitudinal size of the system) for GE and GA at $T$=700\,K.
Because of the specific boundary condition the ripples appear in the
$x$-direction, while in the $y$-direction the fluctuations are much
smaller and for GE they are almost zero.  This leads to a larger
randomness in GA which relates to a less stiff material as compared
to GE~\cite{peeters}. This is illustrated in Fig.~\ref{fig2D} which
shows the corresponding contour plot for GA and GE at $T$=300\,K and
700\,K. Notice the difference in scale for the amplitude.
 For example for GA at 700\,K the
amplitude of the ripples are in the [-4,4]\,\AA~range and
[-2,2]\,\AA~for GE. In the lateral boundaries we see  larger
amplitude variations. The free boundaries in GE exhibit larger free vibrations than GA. This can also be inferred from the top
panel of Fig.~\ref{fig1}(c). Notice that the typical wave lengths ($\lambda$) for the
ripples in GA are very different as compared to the one of GE. Figure~\ref{figFourier} shows the amplitude of
 the Fourier transform of $h_x$,
$\langle|h_q|^2\rangle$ (with an ensemble average  taken over 60
samples),  for GA and GE as function of the wave length along the
$x$-direction (taken in the middle of the system around $y$=0),
$q_x=\frac{2\pi}{\lambda}$ for two different temperatures.
Independent of temperature, the ripples in GA span a wider range of
wave lengths, for example they span
 length scales in the (30-125)\,\AA~and (25-140)\,\AA~ranges, for $T=300$\,K  and $T=700$\,K, respectively.
This compares with ripples in GE having typical length scale around
60\,\AA~for $T=700$\,K~while for $T=300$\,K the membrane is almost flat.

\begin{figure}
\begin{center}
\includegraphics[width=0.8\linewidth]{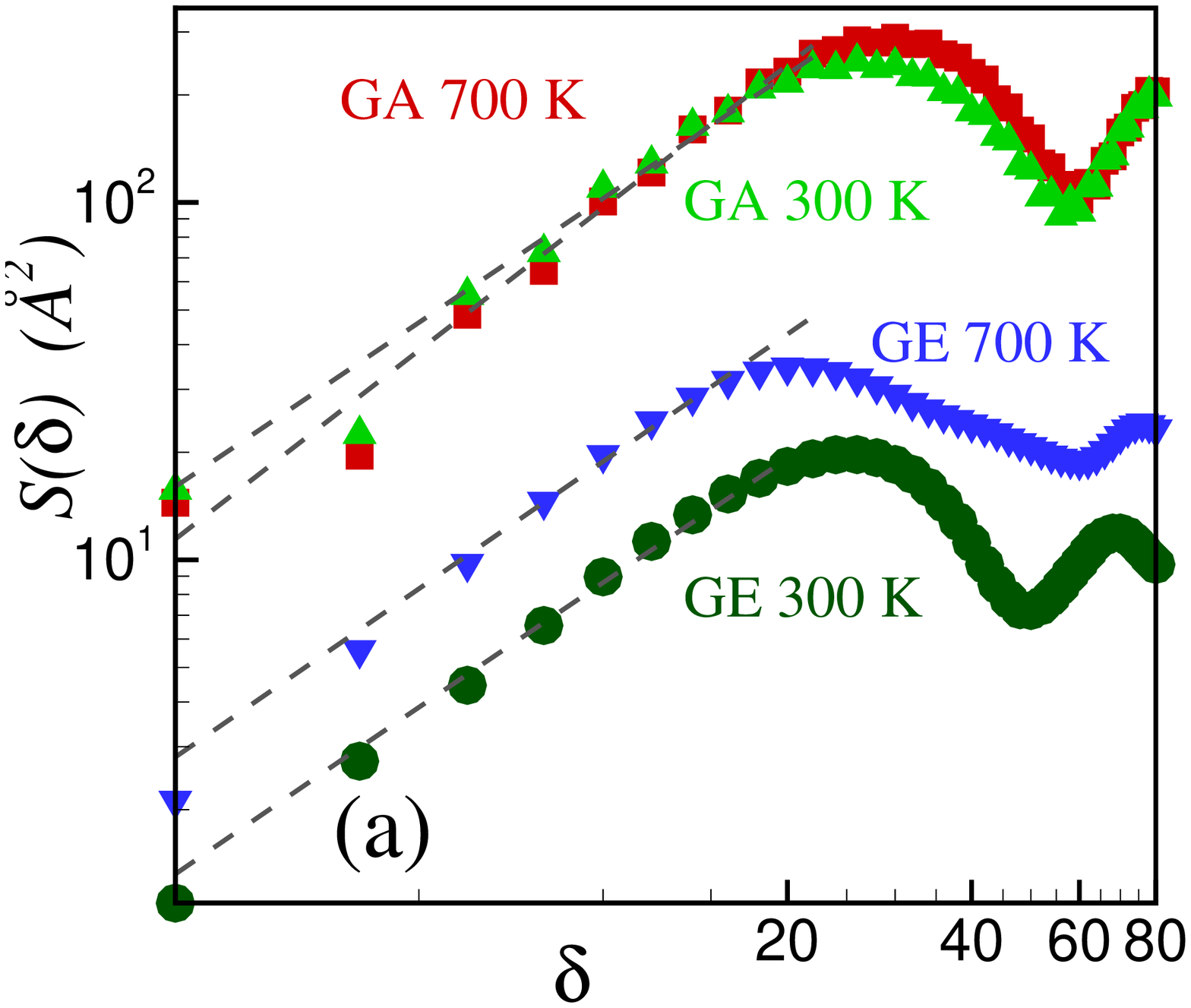}
\includegraphics[width=0.8\linewidth]{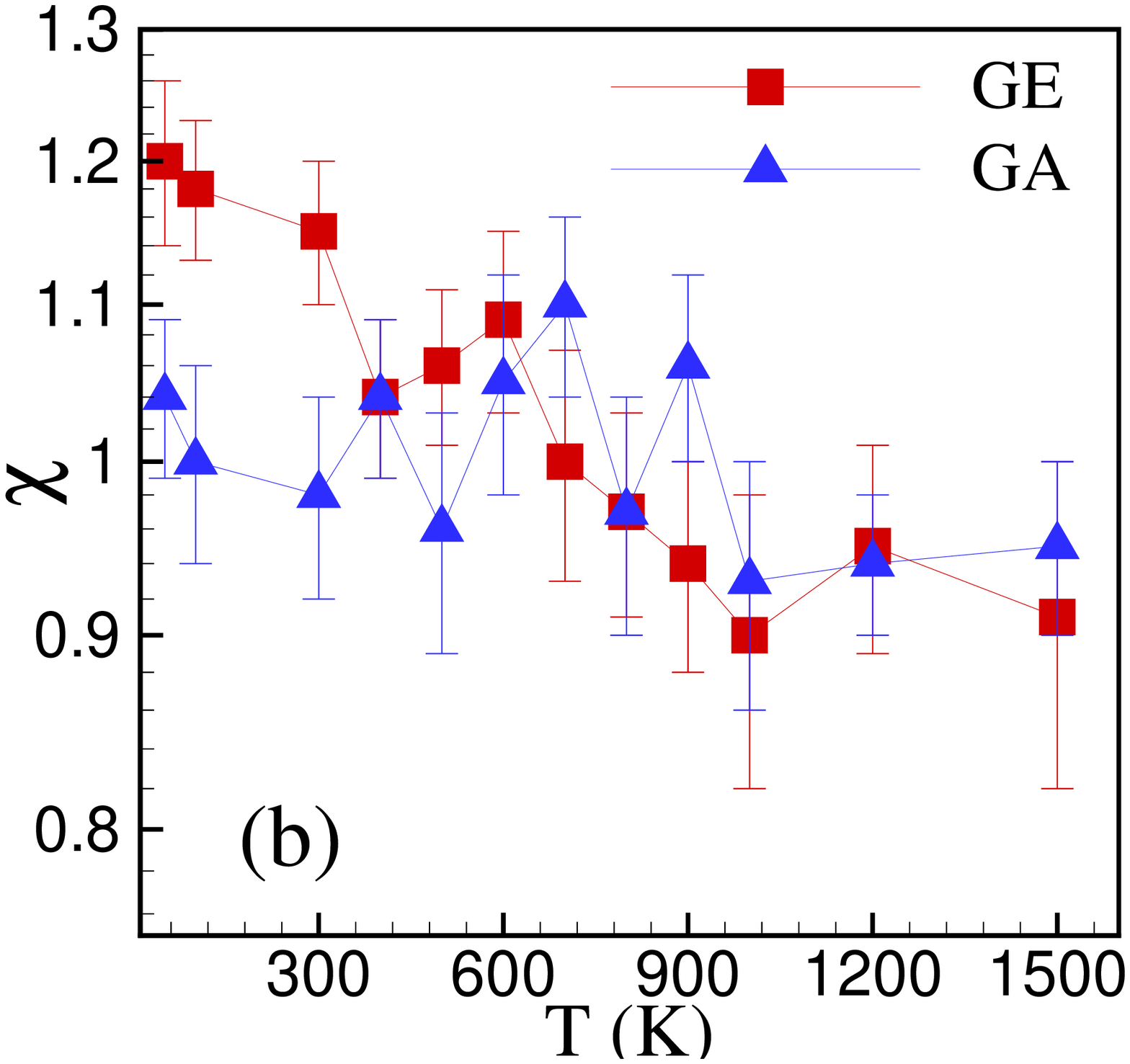}
\caption{(Color online) (a) Variation of the structure function
versus $\delta$ (log-scale) for GE and GA at 300\,K and 700\,K. Here $\delta$ counts the atoms along armchair direction and is about 2\AA.~
Dashed lines are fits  to $\delta^{\chi}$. (b) Roughness exponent as
a function of temperature for GA and GE.  \label{figHH} }
\end{center}
\end{figure}

Next we estimate the roughness exponent $\chi$ of GA (where only the
carbon atoms are considered) which is obtained from the second order
structure function, i.e., $S(\delta)=|\langle
h(\textsl{x}+\delta)-h(\textsl{x}) \rangle|^2$ which scales as
$\delta^\chi$~\cite{neeknanotech} where $\delta$ counts the atoms
along the $x$-direction, e.g. $\delta=4$ refers to the fourth atom
in the armchair direction ($\delta$ is typically 2\,\AA).
Figure~\ref{figHH}(a) shows the variation of $S(\delta)$ versus
$\delta$ for GA and GE for two  temperatures. Notice that
$S(\delta)$ for GA is almost an order of magnitude larger than for
GE indicating the larger corrugation. $S(\delta)$ increases with
$\delta$ up to some critical $\delta$-values, $\delta_c$, which is
related to the above mentioned
 characteristic length scale of the ripples. For $\delta<\delta_c$
the height of the atoms  are correlated  while they become
uncorrelated for $\delta>\delta_c$.  The slope of $S(\delta)$ gives
$\chi$ whose temperature dependence is shown in Fig.~\ref{figHH}(b).
Below room temperature GE has a larger $\chi$ ($\simeq$\,1.2) which
implies that in this temperature range the GE membrane has a
smoother surface as compared to GA. However at high temperatures
both systems approach the situation with random height fluctuations,
i.e. $\chi \sim$\,1. Indeed, one expects that the presence of the
$sp^3$ bonds in GA (making GA effectively a much thicker material
than GE) decreases the roughness with respect to GE while we found
it increases.

\begin{figure}
\begin{center}
\includegraphics[width=0.9\linewidth]{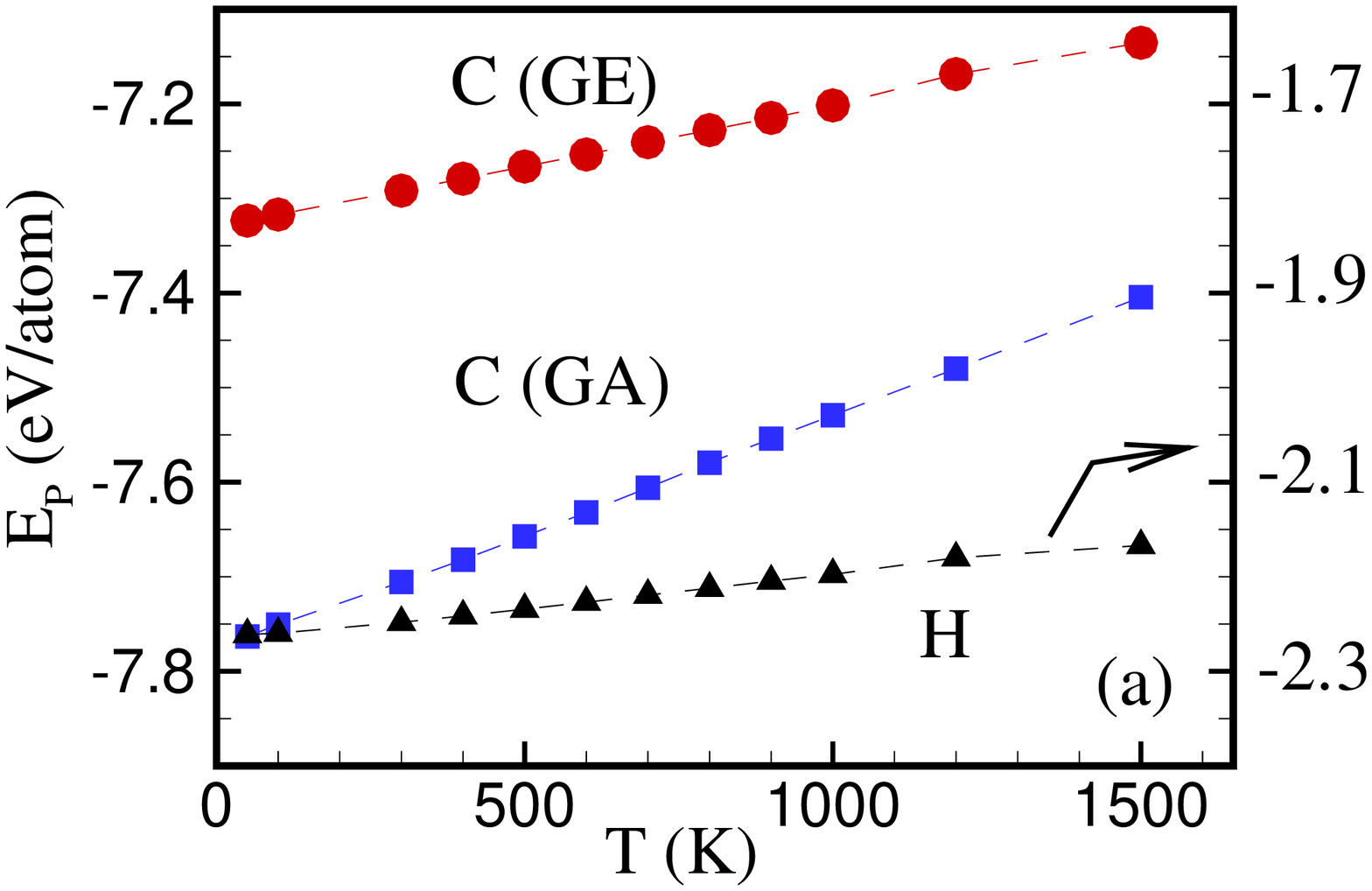}
\includegraphics[width=0.9\linewidth]{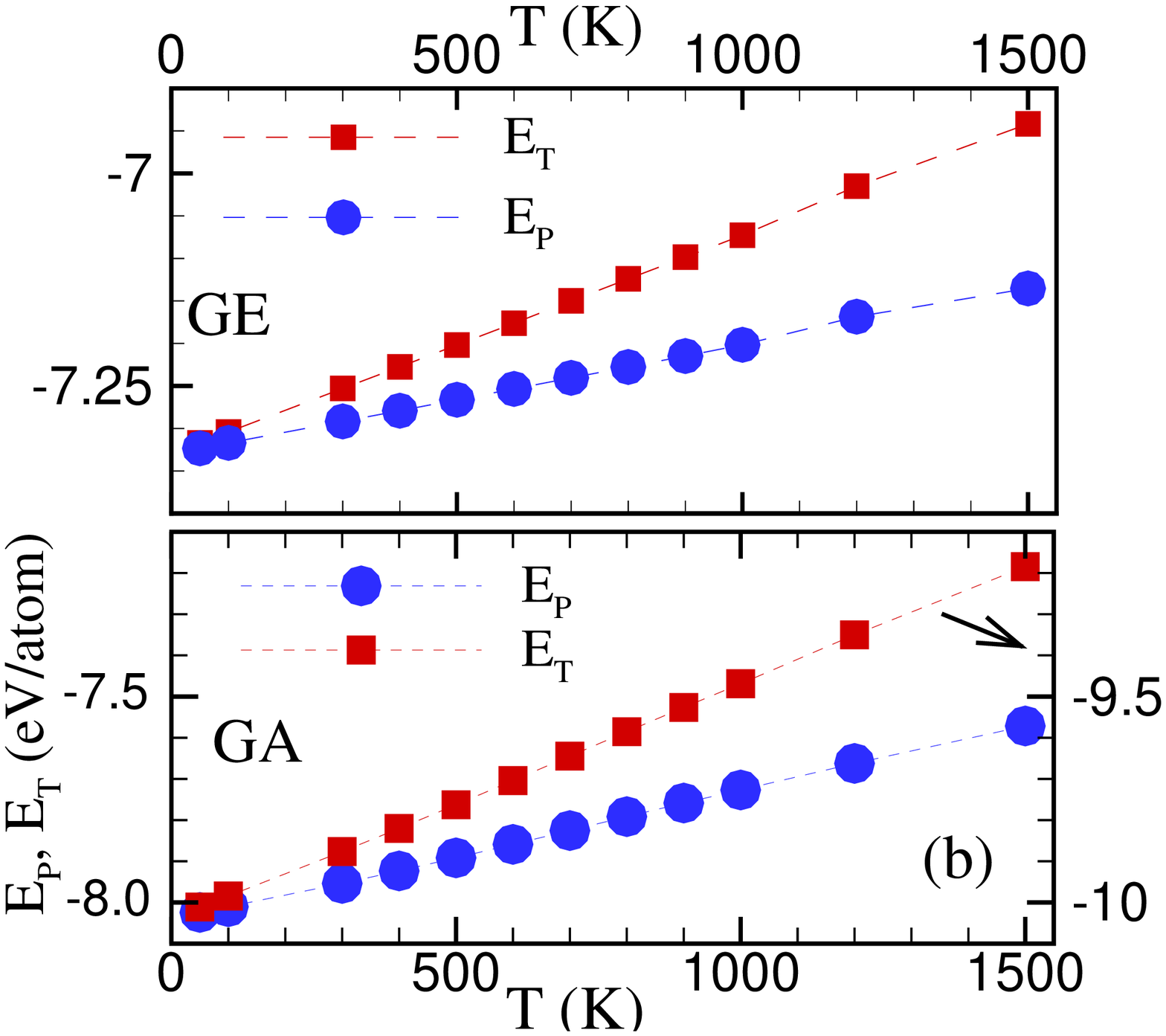}
\caption{(Color online) (a) Variation of the potential energy
($E_P$) per atom versus temperature for both C atoms in GA and GE
and H atoms in GA. The right scale is for H atoms and the left scale
is for C atoms. (b) Variation of total energy ($E_T=E_P+E_K$) per
atom versus temperature for both GA and GE. The right scale is for
GA. The error bars are less than 0.005\,eV/atom.\label{figE} }
\end{center}
\end{figure}

\subsection{Heat capacity}
 Figure~\ref{figE}(a) shows
the variation of the potential energy per atom ($E_P$) for C atoms
in GE, C atoms in GA and H atoms in GA  versus temperature (all data
points were obtained by averaging over an ensemble of 150 samples).
The potential energy is a measure of the binding energy (or bond
energy), i.e. $BE=-E_P$. The binding energy of the C atoms in GA is
larger than the one in GE. The right scale in Fig.~\ref{figE}(a) is
for the potential energy of H atoms in GA. The variation of the
potential energy of H atoms with temperature is smaller than the one
for C atoms in GA. Note that an accurate study of the GA binding
energies at low temperature needs ab-initio molecular dynamic
simulations where quantum fluctuations are included. At low
temperature, where the difference between energy of two quantum
states of the system is larger than the thermal energy, many body
effects and the chemical covalent bond energies become sensitive to
the quantum states. Here we found that the sum of the binding energy
of C-H bonds at 50\,K is 5.012\,eV per atom which is comparable to
the recently reported value at $T$=0\,K, i.e. 5.19\,eV per atom
using DFT calculations~\cite{chBE}.
 The extrapolation of our data to  $T$=0\,K gives 5.02\,eV per atom which indicates
 that quantum corrections are of  order 0.17\,eV per atom or about 3.3$\%$.
Extrapolation in Fig.~\ref{figE}(a), for the binding energy of C
atoms in GE gives 7.33\,eV/atom (for all reported energies the error
bars are less than 0.005\,eV/atom).

 Figure~\ref{figE}(b) shows the total energy per atom (the
sum of potential and kinetic energy of C and H atoms $E_T=E_P+E_K$)
versus temperature (potential energies are also shown in
Fig.~\ref{figE}(b) for comparative purposes. The right scale in
Fig.~\ref{figE}(b) is for the total energy of GA. At a temperature
around 1500\,K we observe  evaporation of H atoms at the free edges.
The total energy varies linearly with
 temperature and gives the corresponding lattice contribution to the molar heat
 capacity at constant volume (the average size of the system after relaxation is taken constant)
$C_V=\frac{dE_T}{dT}$ which for GE is 24.98$\pm$0.14\,J/molK and for
GA is 29.32$\pm$0.23\,J/molK. This is comparable to the proposed
classical heat capacity at constant volume i.e. $C_V=3\Re
\simeq24.94$\,J/molK,  i.e. the Dulong-Petit limit where $\Re$ is
the universal gas constant. The larger heat capacity for GA is due
to the extra storage of vibrational energy  in the C-H bonds. Our
result for GE is in agreement with those obtained from Monte Carlo
simulations of Ref.~\cite{fasolino} but with smaller error bars. The
reported heat capacity for GE in Ref.~\cite{fasolino} is
$C_V\simeq25\pm1.0$\,J/molK for  $T$=1000\,K. The used size for the
GE samples in Ref.~\cite{fasolino} are $13\%$ smaller than those in
the present study.

\section{Conclusion}
 In this paper we studied the
thermal properties of suspended graphane and compared the obtained
results to the one found for graphene. We found that the roughness,
amplitude and wave lengths of the ripples are very different. The
thermal contraction effect for graphane is larger than for graphene.
Above 1500\,K we found that graphane is buckled and starts to loose
H-atoms at the edges of the membrane. Roughness in graphane is
larger than in graphene and the roughness exponent in graphene
decreases versus temperature (from 1.2 to 1.0) while for graphane it
stays around 1.0 implying random uncorrelated roughness. Fourier
analysis of the height of the C atoms showed that the ripples in
graphane exhibit a larger range of length scales as compared to the
one for graphene. Heat capacity of graphane is found to be 14.8$\%$
larger than the one for graphene.

\emph{{\textbf{Acknowledgment}}}.
 This work was supported by the Flemish science foundation (FWO-Vl) and the Belgium Science Policy~(IAP).

\end{document}